\documentclass[useAMS,usenatbib,usegraphicx]{mn2e}
\usepackage[a4paper,centering, totalwidth=520pt, totalheight=700pt]{geometry}
\usepackage{hyperref}
\usepackage{subfig}
\usepackage{graphicx}
\usepackage{color}
\usepackage{pdfpages}
\usepackage{amsmath}
\usepackage{float}

\usepackage[compact]{titlesec}
\titlespacing{\section}{0pt}{*2}{*2}
\titlespacing{\subsection}{0pt}{*2}{*2}
\titlespacing{\subsubsection}{0pt}{*2}{*2}

\newcommand{\subfind}{\textsc{subfind}{ }}
\newcommand{\eagle}{\textsc{eagle}{ }}
\newcommand{\gadget}{\textsc{gadget-2}{ }}

\newcommand{\be}{\begin{equation}}
\newcommand{\ee}{\end{equation}}

\def\ltsima{$\; \buildrel < \over \sim \;$}
\def\simlt{\lower.5ex\hbox{\ltsima}}
\def\gtsima{$\; \buildrel > \over \sim \;$}
\def\simgt{\lower.5ex\hbox{\gtsima}}
\def\la{$\langle$}

\title[Energy equipartition and galaxy sizes]{Energy equipartition between stellar and dark matter particles in cosmological simulations results in spurious growth of galaxy sizes}

\author[Ludlow et al.] {\parbox{18cm}{
    Aaron D. Ludlow$^{1,\star}$,
    Joop Schaye$^{2}$,
    Matthieu Schaller$^{2}$,
    Jack Richings$^{3,4}$
  }\vspace{0.1cm}\\
  $^{1}${International Centre for Radio Astronomy Research, University of Western Australia, 35 Stirling Highway, Crawley,}\\
  {Western Australia, 6009, Australia}\\
  $^{2}${Leiden Observatory, Leiden University, PO Box 9513, 2300 RA Leiden, the Netherlands}\\
  $^{3}${Institute for Computational Cosmology, Department of Physics, University of Durham, South Road, Durham DH1 3LE, UK}\\
  $^{4}${Institute for Particle Physics Phenomenology, Department of Physics, University of Durham, South Road, Durham DH1 3LE, UK}\\
}

\begin{document}
\maketitle
\begin{abstract}
  The impact of 2-body scattering on the innermost density profiles of dark matter haloes is well established.
  We use a suite of cosmological simulations and idealised numerical experiments to show that 2-body scattering is
  exacerbated in situations where there are two species of unequal mass. This is a consequence of mass
  segregation and reflects a flow of kinetic energy from the more to less massive particles.
  This has important implications for the interpretation of galaxy sizes in cosmological hydrodynamic
  simulations, which nearly always model stars with less massive particles than are used for the
  dark matter. We compare idealised models as well as simulations from
  the \eagle project that differ only in the mass resolution of the dark matter component, but keep sub-grid
  physics, baryonic mass resolution and gravitational force softening fixed. If the dark matter particle mass
  exceeds the mass of stellar particles, then galaxy sizes--quantified by their projected half-mass radii,
  ${\rm R_{50}}$--increase systematically with time until ${\rm R_{50}}$ exceeds a small fraction of the
  redshift-dependent mean inter-particle separation, $l$ (${\rm R_{50}}\simgt 0.05\times l$). Our
  conclusions should also apply to simulations that adopt different hydrodynamic solvers, subgrid physics or
  adaptive softening, but in that case may need quantitative revision. Any simulation employing a
  stellar-to-dark matter particle mass ratio greater than unity will escalate spurious energy transfer from
  dark matter to baryons on small scales.
\end{abstract}


\renewcommand{\thefootnote}{\fnsymbol{footnote}}
\footnotetext[1]{E-mail: aaron.ludlow@icrar.org}

\section{Introduction}
\label{SecIntro}

Cosmological simulations of collisionless dark matter (DM) make reliable predictions for the
innermost structure of DM haloes. Such simulations incur relatively 
modest computational cost and have been repeated at ever increasing resolution, exposing the limits of their
reliability \citep[see, e.g.,][]{Stadel2009,Navarro2010}.
Controlling for other numerical parameters--such as time-stepping,
integration accuracy and gravitational softening--their main impediment is 2-body relaxation,
which sets a lower-limit to the spatial resolution of any N-body simulation
(\citealt{Power2003}; hereafter P03; \citealt{Ludlow2018}; hereafter LSB18). This
limitation is well understood and readily accounted for, leading to
widespread agreement on the innermost structure of DM haloes.

Such simulations provide the rudimentary infrastructure for modelling galaxy formation, offering a tangible
connection to observational astrophysics. Current approaches to this problem follow semi-analytic
or halo occupation methods--here the physics of galaxy formation is divorced from the evolution of DM--or
simultaneously model the co-evolution of DM and baryonic fluids.
In both approaches, sub-resolution models for galaxy formation require calibration
against observables before sensible predictions for galaxy populations can be made.
This may overshadow the complex non-linear coupling between
numerical and subgrid parameters, and may mask subtle numerical effects.

One possible issue--which we highlight in this letter--is the importance of 2-body relaxation for the
{\em stellar} component
of simulated galaxies. Stars are treated as collisionless particles in cosmological simulations and, like
DM, their dynamics must be subject to 2-body scattering. Galaxies formed in cosmological simulations, while
calibrated to resemble observed systems, may therefore evolve in a way that is subject to
numerical artefact.

In Section~\ref{SecRelax} we discuss the importance of 2-body scattering in N-body simulations, emphasising
differences between uniform resolution runs and those involving mixtures of DM and stars of
unequal mass, which is the conventional approach. We present simple numerical experiments that
illustrate the effects. In
Section~\ref{SecSimICs} we describe the cosmological simulations used to test the impact of 2-body
scattering on the evolution of stellar systems; their results are presented in Sections~\ref{SSecCollMix} and \ref{SSecSizes}.
We provide some closing remarks in Section~\ref{SecSummary}.

\section{2-body relaxation in an idealised galaxy-halo model}
\label{SecRelax}

Cosmological simulations involve mixtures gas, stars and DM particles typically of unequal mass.
When collisions cannot be ignored, their evolution is subject
to 2-body scattering and, when masses are unequal, to energy equipartition
\citep[e.g.][]{SpitzerHart1971}.
The net energy exchange between species due to these processes can be described by a diffusion equation, with
coefficients that depend on their initial phase-space distributions, and
the ratio of particle masses.

Following \citet{BinneyTremaine2008}, we consider the collisional relaxation time of such a
system, neglecting the
gas component. We define the particle mass ratio, $\mu\equiv m_1/m_2\geq 1$, and the fraction
of mass in $m_2$: $\psi\equiv {\rm M_2/M_1}={\rm N_2}\,m_2/{\rm N_1}\, m_1$, where ${\rm N}_i$ are the
number of particles of species $i$. A test particle that traverses a system of size $R$
will experience $\delta n=\delta n_1 + \delta n_2\approx 2\,\pi\,(\Sigma_1 + \Sigma_2)\, b\,db$ collisions
with impact parameters in the range
$(b,b+db)$, where $\Sigma_1={\rm N_1}/\pi\,R^2$ and $\Sigma_2=\psi\,\mu\,{\rm N_1}/\pi\,R^2$ are the
surface densities of species 1 and 2, respectively. From the impulse approximation, any single encounter
results in a small velocity perturbation ($\delta v\ll v$) perpendicular to the particle's direction
of motion; its trajectory is unaltered. Regardless of its mass, velocity perturbations
are of order $|\delta v_i|\approx 2\,G\,m_i/(b\,v)$ for encounters with particles of mass $m_i$. Such
encounters add incoherently and their cumulative effect will be given by integrating
$\delta v_1^2\delta n_1+\delta v_2^2\delta n_2$ over some range of impact
parameters, $b_{\rm min}$ to $b_{\rm max}$. The relative square velocity change after
traversing the system is given by
\begin{equation}
  \frac{\Delta v^2}{v^2}=\frac{8}{{\rm N_1}}\, \ln\Lambda\, \frac{(1+\psi/\mu)}{(1+\psi)^2},
  \label{eq:delv2}
\end{equation}
where we have assumed a typical velocity $v^2\approx G\,{\rm N_1}\,m_1\,(1+\psi)/R$, and
$\Lambda\equiv b_{\rm max}/b_{\rm min}$ is the Coulomb logarithm. 

For cosmological simulations eq.~\ref{eq:delv2} can be simplified if we identify 
species 1 with DM and species 2 with stars; $\psi$ is then the
stellar-to-DM halo mass ratio, typically $\simlt 0.05$. Assuming equal numbers of baryon and DM
particles $\mu=(\Omega_{\rm M}-\Omega_{\rm bar})/\Omega_{\rm bar}$, where $\Omega_{\rm M}$ and
$\Omega_{\rm bar}$ are the cosmic densities of matter and baryons, respectively. In this case $\mu\geq 1$ and
$\psi \ll 1$, and the ratio of bracketed terms in eq.~\ref{eq:delv2} is close to unity and may be
ignored. If we further assume $b_{\rm max}=R$ and set $b_{\rm min}=b_{90}=G\,(m_1+m_2)/v^2$
as the impact parameter yielding 90$^\circ$ deflections, then
$\Lambda={\rm N_1}\,(1+\psi)/(1+\mu^{-1})\approx {\rm N_1}$ and eq.~\ref{eq:delv2} reduces to
$\Delta v^2/v^2\approx 8\ln{\rm N_1}/{\rm N_1}$. The relaxation of {\em both} species is
driven by encounters with {\em massive} particles. 

The number of orbits a particle must complete so that $\Delta v^2/v^2\approx 1$ defines the
relaxation time, $t_{\rm rel}=t_{\rm orb}/(\Delta v^2/v^2)$. In units of the Hubble time (roughly
the orbital time at the radius\footnote{We define $r_{200}$ as the size
  of a sphere centred on the particle with the minimum potential energy that encloses a mean density
  of $200\times\rho_{\rm crit}$, where $\rho_{\rm crit}(z)=3\,H(z)^2/(8\,\pi\,G)$ is the critical
  density. The corresponding mass is $M_{200}=(800/3)\,\pi\,r_{200}^3\rho_{\rm crit}$
  and circular velocity, $V_{200}=\sqrt{G\,M_{200}/r_{200}}$.}, $r_{200}$),
$t_{\rm H}\approx 2\,\pi\,r_{200}/{\rm V_{200}}$, this can be expressed 
\begin{equation}
  \kappa_{\rm rel}\equiv \frac{t_{\rm rel}}{t_{\rm H}}=\frac{{\rm N_1}}{8\ln {\rm N_1}}\frac{t_{\rm orb}}{t_{\rm H}}=\frac{\sqrt{200}}{8}\frac{{\rm N_1}}{\ln {\rm N_1}}\biggr(\frac{\overline{\rho}\
}{\rho_{\rm crit}}\biggl)^{-1/2},
  \label{eq:trel}
\end{equation}
where $N\equiv N(r)$ is the enclosed particle number, $\overline{\rho}(r)$ the enclosed density,
and $t_{\rm orb}=2\,\pi\,r/V$ is the local orbital time\footnote{A more common definition of
  the relaxation time is based on the number of
  {\em crossings} a particle must execute such that $\Delta v^2/v^2\approx 1$, which differs from our
  definition by a factor $t_{\rm orb}/t_{\rm cross}=\pi$. We adopt the orbital time to define $t_{\rm rel}$
  for consistency with P03: in this case, $t_{\rm orb}/t_{\rm H}\approx (r/V)/(r_{200}/V_{200})$ results in
  eq.~\ref{eq:trel}.} (P03).
When other numerical parameters are chosen wisely, $t_{\rm rel}$ sets a minimum resolved
spatial scale within which collisions cannot be ignored.
The solution to eq.~\ref{eq:trel} thus
defines a ``convergence radius'', $r_{\rm conv}$, which marks the location at
which $\kappa_{\rm rel}\sim 1$ (see, e.g., P03; LSB18).

The value of $\kappa_{\rm rel}$ corresponding to a certain level of convergence must be
obtained empirically by comparing simulations of differing mass
resolution. P03 found that, for DM-only simulations, the circular velocity profile, $V_{\rm c}(r)$,
of an individual Milky Way-mass halo
converges to $\approx 10$ per cent at the radius where $\kappa\approx0.6$; similar convergence in the
{\em average} $V_{\rm c}(r)$ profiles
requires a less conservative value, $\kappa\approx 0.18$, regardless of halo mass (LSB18).
A convenient approximation is given by
$r_{\rm conv}=0.174\, \kappa_{\rm rel}^{2/3}\,l$, where $l={\rm L_{box}}/N_{\rm part}^{1/3}$ is the
mean inter-particle spacing in physical units, and $\kappa_{\rm rel}=0.18$ (LSB18).

When $\mu\ne 1$, two-body collisions also lead to a {\em segregation} of the two
components: massive particles will, on average, lose energy to less
massive ones, causing them to congregate in halo centres while heating the low-mass component.
This “mass segregation” signals the onset of energy equipartition.

The simple 2-component toy model of \citet{Spitzer1969} suggests that the segregation timescale, $t_{\rm seg}$, is
shorter than $t_{\rm rel}$ by a factor roughly equal to the ratio of the particle masses:
\begin{equation}
  t_{\rm seg}=\frac{t_{\rm rel}}{\mu}\approx\frac{{\rm N_1}}{8\,\ln {\rm N_1}}\,\frac{t_{\rm orb}}{\mu},
  \label{eq:tseg}
\end{equation}
Homogeneous mixtures of particles of different mass will therefore segregate at radii
$r_{\rm seg} \geq r_{\rm conv}$ provided $\mu \geq 1$. A simple estimate of
$r_{\rm seg}$ therefore follows from eq. \ref{eq:trel} (or from
$r_{\rm conv}=0.174\, \kappa_{\rm rel}^{2/3}\,l$) if $\kappa_{\rm rel}$ is replaced by
$\kappa_{\rm seg}=\mu\,\kappa_{\rm rel}$. Whether equipartition can be reached, however,
depends on the ratios of particle mass, $\mu$, and of the total mass of each component,
${\rm M_1/M_2}$, and $r_{\rm seg}$ should therefore be viewed as an upper limit.
(Simple analytic estimates and numerical results suggest that
full equipartition may not be possible if ${\rm M_1}\simgt {\rm M_2}\,\mu^{-3/2}$, which is
almost always the case in DM-dominated galaxies.)

\begin{figure}
  \vspace{-0.2cm}
  \includegraphics[width=0.49\textwidth]{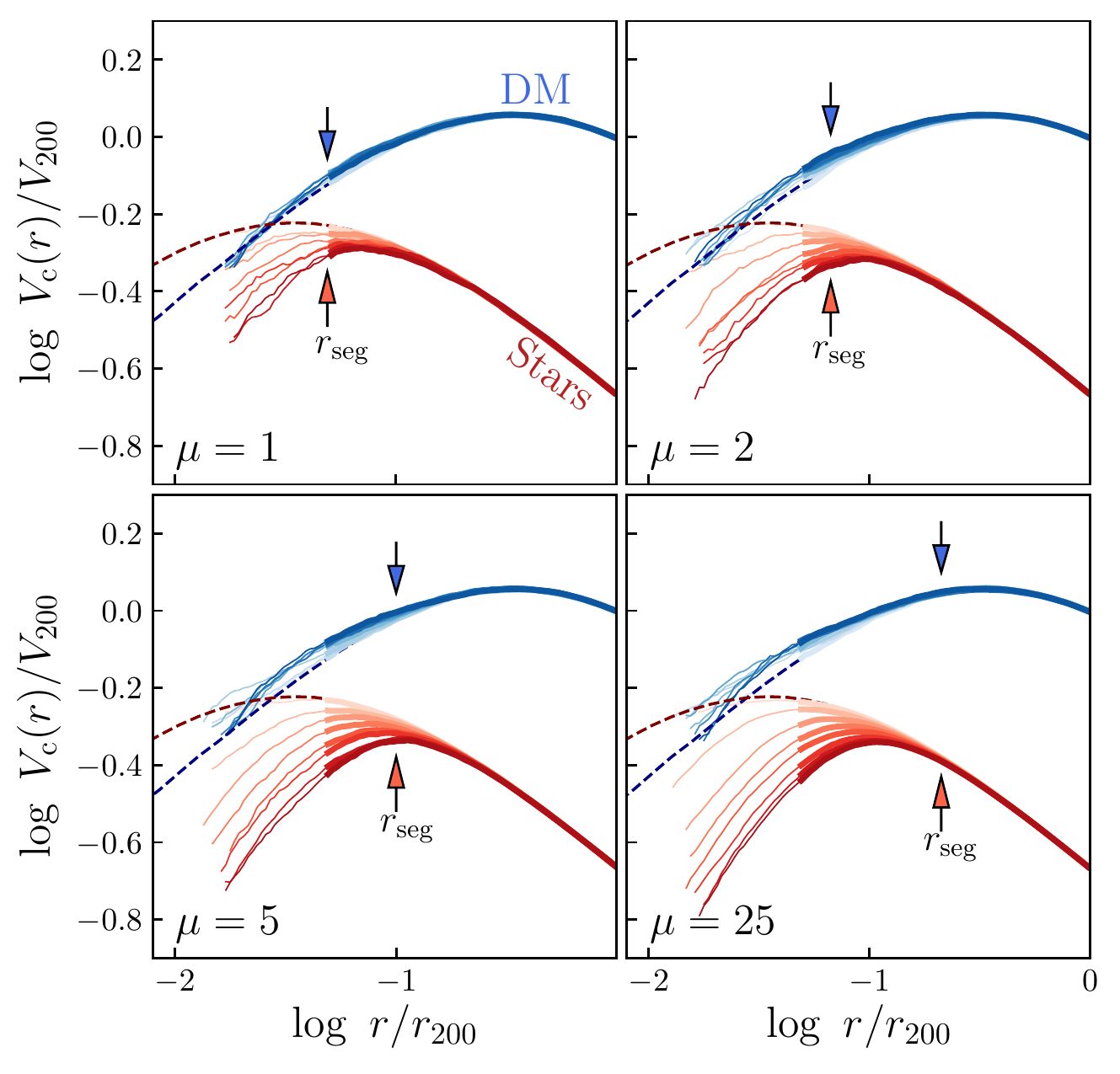}
  \caption{Circular velocity profiles of DM (blue lines) and stars
    (red lines) in a set of idealised numerical simulations starting
    from equilibrium initial conditions (dashed curves). The DM halo
    is sampled with $N_{1}=5\times 10^4$ particles;
    the stellar component, also modelled using
    collisionless particles, has a mass fraction of 2.7 per cent of the system's
    total mass, but a total number of particles proportional to
    ${\rm N_2}\propto\mu\, {\rm N_1}$, where $\mu=1$, 2, 5 and 25 (top to bottom,
    left to right). Different tints and shades correspond to earlier and later
    outputs of the simulation, respectively, which are spaced linearly from $t=0$
    to $t\approx 13.3$ Gyr. For individual profiles,
    the thick lines extend to the convergence radius dictated by 2-body relaxation
    (eq.~\ref{eq:trel}, with $\kappa_{\rm rel}=0.6$), and arrows mark the radius
    $r_{\rm seg}$ (eq.~\ref{eq:tseg}).}
  \label{fig1}
  \vspace{-0.25cm}
\end{figure}

As $\mu\rightarrow 1$, the importance of mass segregation
diminishes. Nevertheless, different species may still structurally evolve
through 2-body scattering and, as we show below, this evolution is sensitive to the
{\em initial} segregation of each component (The sensitivity to initial
  segregation becomes clear when comparing the idealised tests presented in
  Figure~\ref{fig1} to the collisionless cosmological runs in Figure~\ref{fig2}, which start
  with both species equally mixed.).

Figure~\ref{fig1} shows results from numerical experiments designed to illustrate these effects.
We consider here idealised equilibrium systems composed of a galaxy embedded within a DM halo.
Both are modelled as spherical, collisionless \citealt{Hernquist1990} spheres with 
galaxy-to-halo mass ratio $M_{\rm gal}/M_{\rm h}=0.027$ (close to the ``peak'' galaxy formation efficiency
of \citealt{Behroozi2013})
and ratio of scale radii $r_{\rm half}/a_{\rm h}=0.25$ ($r_{\rm half}$ and $a_{\rm h}$ are the
galaxy half-mass radius and halo scale radius, respectively). Initial conditions, constructed 
using GalIC \citep{Yurin2014}, differ only in the stellar-to-DM particle mass ratio, $\mu$.
GalIC assigns particle phase-space coordinates iteratively in order to optimally
match an underlying analytic distribution function, which is itself a
  stable solution to the collisionless Boltzmann equation and therefore in collisionless
  equilibrium\footnote{We have verified that GalICs yields stable particle configurations
    for our experiments by carrying out tests for which $\mu=1$ (to eliminate mass segregation)
    and ${\rm N_1}=10^6$ (to suppress 2-body scattering). Our tests confirm that the mass profiles of
    species 1 and 2 remain stable at $r\simgt r_{\rm conv}$ for a Hubble time.}. We adopt
${\rm N_1}=5\times 10^4$ (for DM), and consider $\mu=1$, 2, 5 and 25. All runs used the same softening length,
$\epsilon/l_h=0.1$ ($l_h=[3/4\,\pi\,N_1]^{1/3}$ is the  Wigner-Seitz radius), and were evolved using \gadget
\citep{Springel2005b} for $t\approx 13.3$ Gyr. (We have verified that our simulation results
  are robust to small changes in timestepping and softening length.)
Because these systems are initially in collisionless equilibrium, any evolution away
from the initial state must be driven by 2-body scattering.

Different panels correspond to
different $\mu$, as indicated. Solid blue curves show $V_{c,1}(r)$ for the DM,
and solid red curves show $V_{c,2}(r)$ for stars; tints and shades encode
the time evolution, which increases linearly from $t=0$ (light) to $t\approx 13.3\,{\rm Gyr}$
(dark). Dashed lines of corresponding colour show the initial profiles used to construct
the galaxy/halo models. For each curve (except the initial profiles) thick lines extend
down to the convergence radius expected from eq.~\ref{eq:trel} (for $\kappa_{\rm rel}=0.6$);
thin lines extend to the radius enclosing 100 DM particles. 

DM profiles are reasonably stable for $r\simgt r_{\rm conv}$, as is the case for stars if 
$\mu=1$. Note, however, that as $\mu$ increases, the curves deviate systematically
from their initial profile at radii $\simgt r_{\rm conv}$; this is particularly true for
stars. The arrows mark $r_{\rm seg}$ calculated from eq.~\ref{eq:trel} after replacing $t_{\rm rel}$
by $t_{\rm seg} =t_{\rm rel}/\mu$ . For $\mu\simlt 5$, these arrows track more closely the radii at
which $V_c(r)$ profiles first show noticeable differences from their initial values. Note also that
the segregation of the stars and DM is much more prominent when $\mu$ is large: DM haloes develop
denser centres while the stellar component gradually expands. Importantly,
even for $\mu=1$ there is considerable evolution in ${\rm V_{c,2}}(r)$ for $r\simlt r_{\rm conv}$.
This is because 2-body collisions will tend to homogenise populations that are initially
segregated. 2-body scattering can be thought of as a diffusion process with different coefficients
describing the first- and second-order processes. For particular initial configurations--which depend
on $\mu$, $M_1/M_2$ and the spatial segregation of each species--energy equipartition may give rise
to mass segregation. For the particular case of $\mu=1$ the diffusion coefficients are equal, and
2-body scattering will lead to a {\em mixing} of the two components. In that case, a centrally compact
stellar component will tend to become more diffuse with time, as seen in the upper left panel of
Figure~\ref{fig1}, even if it was {\em constructed} to be in collisionless
equilibrium initially (note that scattering-driven diffusion is largely
confined to within the convergence radius, as expected for $\mu=1$).

\renewcommand{\tabcolsep}{4pt}
\begin{center}
 \begin{table}
   \caption{Basic numerical parameters used for our cosmological simulations. ${\rm N_1}$ refers
     to the number of ``dark matter'' particles of species 1; ${\rm N_2}$ to the number of ``gas''
     particles, species 2. The corresponding particles masses are $m_1$ and $m_2$, respectively,
     and their ratio is denoted $\mu\equiv m_1/m_2$; ${\rm L_{box}}$ is the simulation box size.
     The run type is also provided: DMO refers to runs for which both species were assumed to be
     collisionless particles, and ``\textsc{eagle}'' refers to hydrodynamical runs carried out
     using the Reference model of the \eagle project.}
   \begin{tabular}{c c r r c c c c c}\hline \hline
     &  ${\rm N_1}$     & ${\rm N_2}$ &        $m_1$            &     $m_2$              & $\mu$       &  ${\rm L_{box}}$ &  Type &\\
     &                  &             & [$10^5\, {\rm M}_\odot$]&[$10^5\, {\rm M}_\odot$]&             &   [Mpc]          &       &\\\hline
     & 752$^3$          &     0       &         1.8             &        0               &  --         &  12.5 &  DMO &\\
     & 188$^3$          &  188$^3$    &         97.0            &       18.1             &  5.36       &  12.5 &  DMO &\\
     & 7$\times$188$^3$ &  188$^3$    &         13.9            &       18.1             &  0.77       &  12.5 &  DMO &\\
     & 376$^3$          &  376$^3$    &         97.0            &       18.1             &  5.36       &  25.0 &  \eagle &\\\vspace{0.25cm}
     & 7$\times$376$^3$ &  376$^3$    &         13.9            &       18.1             &  0.77       &  25.0 &  \eagle &\\
   \end{tabular}
   \label{TabSimParam}
 \end{table}
\end{center}

\section{Cosmological Simulations}
\label{SecSims}

\subsection{Simulation set-up}
\label{SecSimICs}

DM haloes and their associated galaxies form hierarchically through accretion and mergers and are,
at best, {\em quasi-}equilibrium structures. It is therefore worthwhile to test the importance of
mass segregation and 2-body scattering in cosmological simulations that include two particle species.
The remainder of the paper will focus on such simulations.

All cosmological runs used parameters consistent with the \citet{Planck2014} data release:
$h\equiv H_0/(100\, {\rm km\, s^{-1}\, Mpc^{-1}})=0.6777$ is the Hubble parameter;
$\sigma_8=0.8288$ the ($z=0$) rms density fluctuation in 8 $h^{-1}{\rm Mpc}$ spheres;
and $\Omega_{\rm M}=1-\Omega_\Lambda=0.307$ and $\Omega_{\rm bar}=\Omega_{\rm M}-\Omega_{\rm DM}=0.04825$,
are the energy density parameters in units of $\rho_{\rm crit}$. 

\begin{figure}
  \vspace{-0.2cm}
  \includegraphics[width=0.49\textwidth]{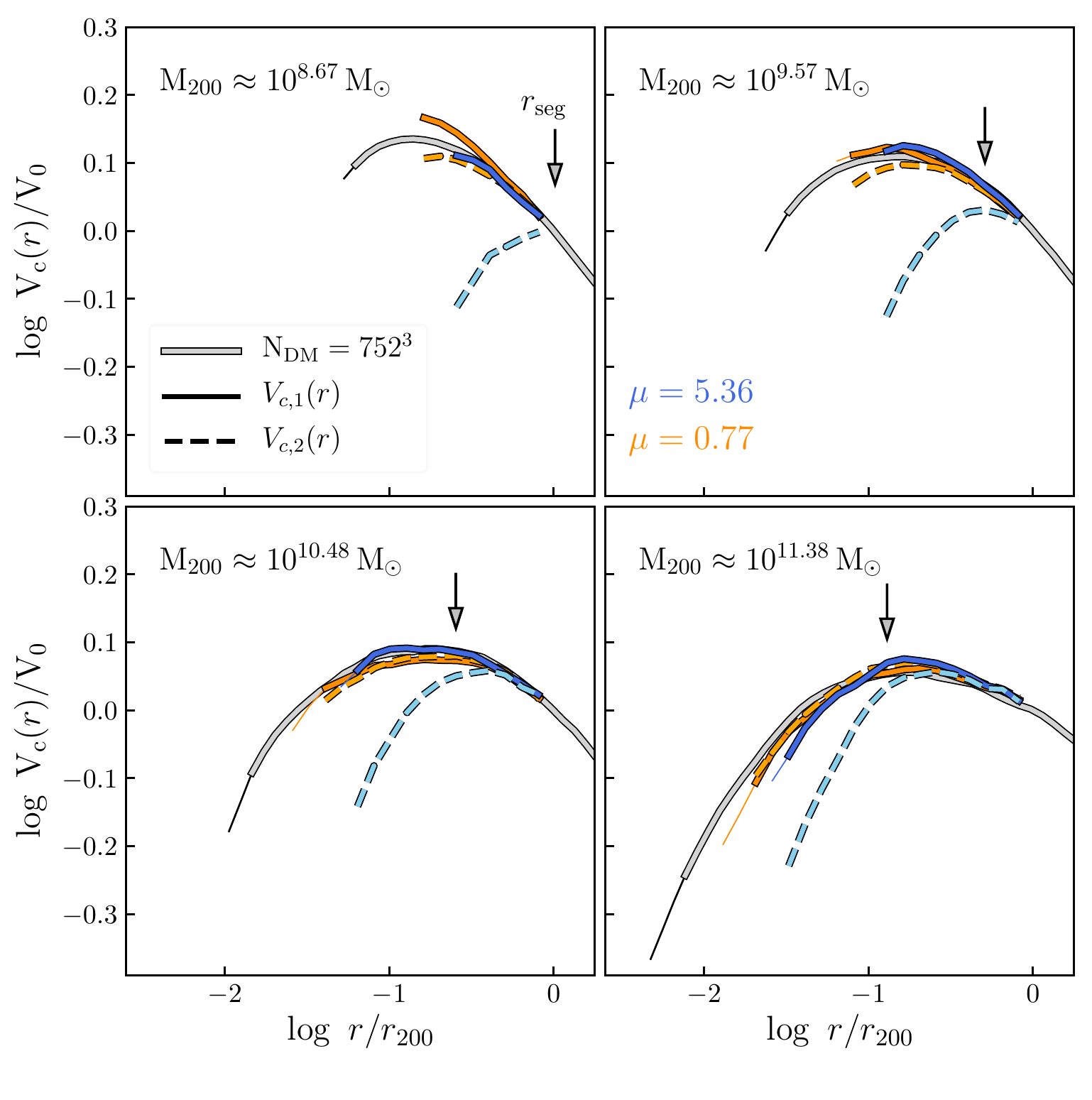}
  \caption{Median ($z=0$) circular velocity profiles of DM haloes in 
    simulations of collisionless particle mixtures. Different panels
    correspond to different masses, ${\rm M_{200}}$, which increase from
    ${\rm M}_{200}=10^{8.5}\, {\rm M_\odot}$ by successive factors
    of 8 between panels. Blue curves correspond to the run with
    ${\rm N_{DM}=N_{gas}}=188^3$ ($\mu=5.36$); orange curves to the one
    with ${\rm N_{gas}=188^3}$ and ${\rm N_{ DM}=7\times 188^3}$ ($\mu=0.77$).
    Solid curves represent DM particles whereas dashed curves represent
    ``gas'' particles. Grey curves correspond to the median
    $V_c(r)$ profiles of haloes in our single-component DM-only run carried
    out with ${\rm N_{part}=752^3}$ particles. For all profiles we use thick line
    segments for $r>0.055\,l$ and thin lines extend to the P03
    convergence radius ($\kappa=0.6$). Downward pointing arrows denote the radius
    $r_{\rm seg}=0.055\,\mu^{2/3}\, l$, below which we expect substantial
    segregation of DM and ``gas'' particles in the $\mu=5.36$ run.}
  \label{fig2}
\end{figure}

We use three cosmological simulations that echo those used by \citet{BinneyKnebe2002} to
investigate 2-body scattering in cosmological DM-only simulations. 
The first evolves the DM with ${\rm N_{part} = 752^3}$ equal-mass
particles ($m_{\rm DM}=1.8\times 10^5\,{\rm M_\odot}$).
The second uses two particle
species of equal {\em abundance}, ${\rm N_1=N_2=188^3}$, but with a mass ratio
$\mu=\Omega_{\rm DM}/\Omega_{\rm bar}\approx 5.36$; this run is analogous
to most cosmological hydrodynamical simulations (DM and baryons are sampled with
equal particle numbers) but differs by modelling both species as collisionless fluids.
The masses of DM  (species 1) and ``gas'' particles (species 2)
are, respectively,
$m_{\rm DM}=97.0\times 10^5\,{\rm M_\odot}$ and $m_{\rm gas}=18.1\times 10^5\,{\rm M_\odot}$. 
The final run also adopts a two collisionless components, but with unequal particle numbers:
${\rm N_1}/7={\rm N_2}=188^3$ and hence $\mu=(1/7)\,\Omega_{\rm DM}/\Omega_{\rm bar}\approx 0.77$
(or $m_{\rm DM}=13.9\times 10^5\,{\rm M_\odot}$). Initial conditions were created from
those described above by splitting DM particles into seven equal-mass particles and arranging them
on a cubic lattice in a manner that preserves a force-free unperturbed particle load.
All runs used a linear box size of $L_{\rm box}=12.5$ Mpc (comoving) and 
identical phases and amplitudes for mutually resolved modes. They differ only in the number of particles
of species 1, and hence in $\mu$.

The particle mixture models were repeated for the second set of simulations
(in a larger volume; ${\rm L_{box}}=25$ Mpc) but with species 2
treated as a gaseous fluid ($N_{\rm gas}=N_2=376^3$).
These runs employ cooling, star formation and feedback from stars and active galactic nuclei
in accord with the Reference model of the \eagle program \citep[see][]{Schaye2015}.
They differ only in the number of DM particles: one has $N_{\rm DM}=N_{\rm gas}=376^3$
($\mu\approx 5.36$) and the other $N_{\rm DM}=7\times N_{\rm gas}=7\times 376^3$ ($\mu=0.77$).
At $z=0$, star particles have an {\em average} mass of roughly 65 percent of the
primordial gas particle mass (stars lose mass to gas particles through stellar winds as they evolve).
As a result, $\mu_\star\equiv m_{\rm DM}/\langle m_\star\rangle \approx 1.18$ in our runs, but we neglect this small
departure from unity.

All runs used the same softening length for both species, which is a fixed fraction of the
mean {\em baryonic} inter-particle separation\footnote{For $N_{\rm p}=752^3$,
  we use the DM inter-particle spacing.}:
$\epsilon/l=0.04$ (comoving) for $z>2.8$, and
$\epsilon/l=0.01$ (physical) thereafter. Haloes were identified using \subfind
\citep{Springel2001b}, which returns the coordinate of the particle with the
minimum potential energy, $\mathbf{x}_{\rm MB}$, as well as $r_{200}$,
${\rm M_{200}}$ and ${\rm V_{200}}$.

As in Figure~\ref{fig1}, we focus our analysis on the circular velocity profiles of each mass
component, and use subscripts to denote the relevant species. (For example, $V_{c,1}(r)$
refers to the circular velocity profile of particles of mass $m_1$.)
Hereafter, for clarity, we drop explicit reference to DM or
baryonic particles, even in the hydrodynamic runs, but instead identify DM with species 1 and stars
with species 2. We do not consider the mass profiles of gas particles in our hydrodynamic simulations.

\subsection{Cosmological simulations with unequal-mass collisionless particles}
\label{SSecCollMix}

Figure~\ref{fig2} shows the median ($z=0$) circular velocity profiles of haloes in four separate mass bins in our collisionless cosmological
runs. Grey curves correspond to the uniform resolution (${\rm N_{part}}=752^3$) simulation, which can be used to assess
convergence in the lower-resolution runs. Blue curves correspond to the run with $N_1=N_2=188^3$ ($\mu=5.36$); orange
to the one with $N_1/7=N_2=188^3$ ($\mu=0.77$). Thick lines extend down to $r_{\rm conv}=0.055\, l$ (LSB18),
where $l$ is the mean inter-particle spacing\footnote{If we were to use instead
  $l={\rm L_{box}/N_{tot}}$ , where ${\rm N_{tot}=N_1+N_2}$, $r_{\rm conv}$ would be smaller by a factor of
  $2^{1/3}\approx 1.26$. This will not affect the interpretation of our results, so we opt for the more
  conservative estimate of $r_{\rm conv}$.}
of particles of mass $m_1$; thin lines to the $r_{\rm conv}$
expected from eq.~\ref{eq:trel} with $\kappa_{\rm rel}=0.6$. To aid the comparison, all curves
have been normalised to $V_0=\sqrt{G\, M_{200,i}/r_{200}}$, where $M_{200,i}$ is the mass of species $i$ enclosed
by $r_{200}$.

This figure prompts a few comments. First, notice that, for simulations involving particle mixtures, the $V_{c,1}(r)$
profiles agree reasonably well with those of equal-mass haloes in the uniform resolution run 
(the solid coloured curves align closely with the grey curves). The largest differences in
$V_c(r)$ are $\simlt 10$ per cent for $r > r_{\rm conv}$, as expected. Particles of mass $m_2$, however, behave
differently depending on $\mu$. For $\mu\approx 0.77$ (dashed orange lines), the circular velocity profiles of
species 1 and 2 are similar: both deviate by $\simlt 10$ per cent from the high-resolution run for all
$r > r_{\rm conv}$ and all halo masses considered. This is expected: since $\mu$ is close to 1,
and both species are initially well-mixed, it follows that
$t_{\rm seg}\approx t_{\rm rel}$ and $r_{\rm seg}\approx r_{\rm conv}$, and both species should remain approximately
homogeneous at $r\simgt r_{\rm conv}$ at all times. For $\mu\approx 5.36$, however, this is not the case:
for $r\simlt r_{\rm seg}$, $V_{\rm c,2}(r)$ is considerably
lower than what is expected for a purely collisionless system, consistent with mass segregation driven by
2-body scattering (this confirms the results of \citealt{BinneyKnebe2002}).
The radius below which this suppression becomes significant coincides roughly with $r_{\rm seg}$
(downward pointing arrows), approximated by $r_{\rm seg}=0.174\,(\mu\,\kappa_{\rm rel})^{2/3}\,l\approx 3.1\, r_{\rm conv}$
(assuming $\kappa_{\rm rel}=0.188$; LSB18).

\subsection{Impact of 2-body scattering on galaxy sizes}
\label{SSecSizes}

Many cosmological simulations spawn one star particle
per gas particle, which typically have comparable masses but are
$\approx \Omega_{\rm bar}/(\Omega_{\rm M}-\Omega_{\rm bar})$
times less massive than the DM particles.
Other simulations attempt to increase the resolution of the
stellar component by generating multiple star particles per gas particle
which are considerably less massive \citep[e.g.][]{Dubois2014,Revaz2018}. 
Galaxies formed in both types of simulations may be subject to equipartition
effects, which may have important implications for the
interpretation of galaxy sizes, among other properties.

What impact does equipartition have on galaxy sizes in cosmological
hydrodynamical simulations?
Figure~\ref{fig3} summarises the results of our tests. Each panel shows
the median projected half-stellar mass radii, ${\rm R_{50}}$, as a function
of galaxy stellar mass (masses are defined using bound stellar
particles within a 100 physical kpc aperture centred on $\mathbf{x}_{\rm MB}$)
at four different redshifts: $z=0$, 0.5, 1 and 2. We
use blue curves for $\mu=5.36$ and orange curves for $\mu=0.77$.
The vertical dashed lines correspond to 100 primordial gas
particles, dotted lines to 2000. These runs use identical baryonic
mass resolution, force softening (arrows indicate $2.8\times\epsilon$)
and sub-grid physics models; they differ {\em only} in DM particle mass.

Galaxy sizes show clear differences between these runs,
both in their mass and redshift dependence. Consider first $z=0$
(upper-left panel). For $\mu=5.36$, the median size-mass relation
flattens abruptly for stellar masses ${\rm M}_\star\simlt 2000\, m_{\rm gas}$ (dotted vertical line)
below which ${\rm R}_{50}\approx 2.8\,{\rm kpc}$, regardless of ${\rm M_{\star}}$.
For $\mu\approx 0.77$ this is not observed: sizes
continue to decrease monotonically with decreasing ${\rm M}_\star$ to the
lowest mass-scale considered ($\approx 10$ stellar particles).
Similar results are seen at $z=0.5$ for $\mu=5.36$, although in this case ${\rm R}_{50}$ levels-off
at lower mass (${\rm M}_\star\approx 10^{8.7}\,{\rm M_\odot}$), and correspondingly smaller size
(${\rm R}_{50}\approx 2\,{\rm kpc}$). For $\mu=0.77$ galaxy sizes evolve very little from
$z=0.5$ to $z=0$ (thin lines, repeated in all panels, show the $z=0$ size-mass relations
for comparison).

Note as well that, for the different $\mu$ values, sizes begin to converge at higher
redshift: by $z=2$, for example, they are virtually indistinguishable for galaxies
resolved with more than $\approx 100$ particles. Intriguingly, convergence is attained
at all $z$ provided sizes exceed the physical convergence radius of haloes in the
$\mu=5.36$ run (shown here as $r_{\rm conv}=0.055\,l$ and highlighted using a red
horizontal line; see LSB18).

Although using $\mu\approx 1$ will minimise the spurious transport of energy
between particle species, we emphasise that by itself it does not
guarantee that the simulations are immune to numerical effects.
Convergence tests that simultaneously increase both the DM and baryonic
resolution, and use $\mu \approx 1$, are required to test in which regime
the results are robust.


\section{Summary and Discussion}
\label{SecSummary}

\begin{figure}
  \vspace{-0.2cm}
  \includegraphics[width=0.49\textwidth]{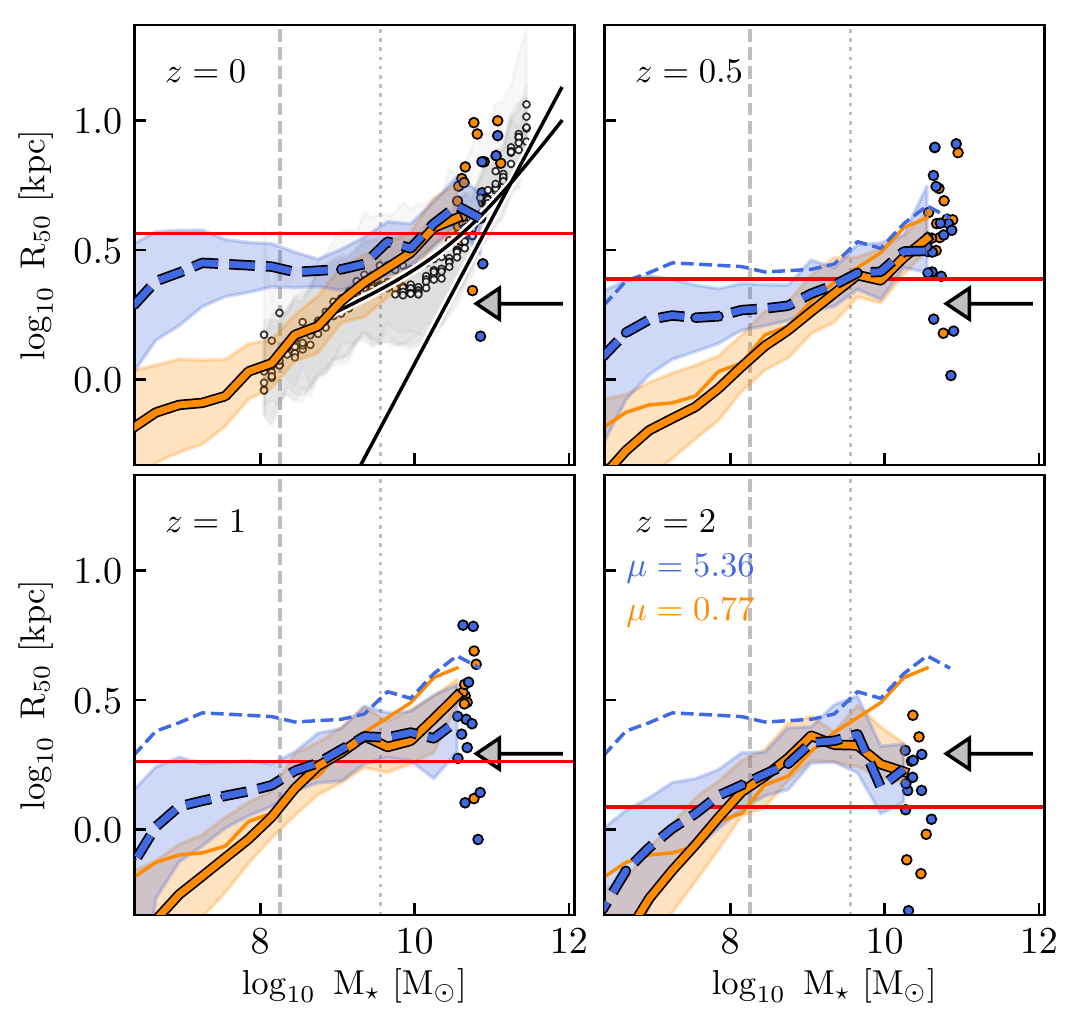}
  \caption{Projected half-mass radius as a function of stellar mass at $z=0$,
    0.5, 1 and 2. Dashed (blue) lines correspond to $\mu=m_{\rm DM}/m_{\rm gas}=5.36$
    and solid (orange) lines to $\mu=0.77$. Thin lines, repeated in all panels,
    show the $z = 0$ relations for comparison. The vertical dashed and dotted lines
    indicate the mass scales of 100 and $2000\times m_{\rm gas}$, respectively; 
    red lines mark $r_{\rm conv}$ for the DM component of the $\mu=5.36$ run;
    arrows mark the spline softening lengths, $2.8\times \epsilon$, above which
    gravitational forces are exactly Newtonian. The much stronger evolution of ${\rm R_{50}}$
    when $\mu=5.36$ is due to numerical mass segregation. For comparison, we plot
    circularized half-light radii for early- and late-type galaxies in SDSS
    (thick black line; \citealt{Shen2003}) and GAMA (points; \citealt{Lange2015}).}
  \label{fig3}
  \vspace{-0.5cm}
\end{figure}

Previous studies of galaxy sizes in cosmological simulations report trends similar to
those in Figure~\ref{fig3} for $\mu=5.36$. In {\textsc{eagle}}, \citet{Furlong2017} note that galaxy
sizes increase systematically with increasing ${\rm M_\star}$  and with decreasing redshift.
They identified a sample of passive galaxies between $z=1.5$ and 2
that remain quiescent centrals until $z=0$: all grow in size between their identification
redshift and $z=0$. They report that compact centrals at $z=2$ 
grow secularly by ``stellar migration'' to the present day.

\citet{Campbell2017} present convergence tests of projected half-mass radii in the Apostle simulations
\citep[][$\mu\approx 5.36$]{Sawala2016}. Comparing low-, intermediate- and high-resolution runs they show that
${\rm R_{50}}$ flattens at a characteristic scale comparable to the
spline softening length.
Our results indicate that sizes are subject to spurious growth below scales comparable to the
{\em convergence} radius ($\approx 0.055\,l$) which are close to those
quoted by \citet{Campbell2017}. We can distinguish between softening and 2-body
scattering as the culprit for this resolution dependence using the 
redshift evolution of ${\rm R_{50}}$. If softening is the cause, then ${\rm R_{50}}\approx \epsilon$ should set the minimum
size {\em at all redshifts}, whereas 2-body scattering would give rise to a {\em slow growth} 
of ${\rm R_{50}}$ for poorly-resolved galaxies. Our results support the latter explanation (Figure~\ref{fig3}). 

Similar results were recently reported for galaxy sizes in the Illustris TNG50 simulation \citep{Pillepich2019}.
In TNG50, for which $\mu\approx 5.3$, the sizes of low-mass
galaxies flatten at systematically
larger physical scales, and at higher stellar masses, as mass-resolution decreases. These results are {\em not} consistent
with softening setting a minimum physical size to low-mass galaxies. In TNG100, \citet{Genel2018} also report
a flattening of sizes for low-mass galaxies, an effect that becomes more pronounced for quiescent systems.
They also note that, during quenched phases, galaxy sizes increase systematically with time,
particularly among poorer-resolved
low-mass systems, despite little growth in stellar mass over the same period. The secular growth of sizes
of non-star forming galaxies (e.g. dwarfs or ellipticals) is an {\em expected consequence} of (spurious)
energy equipartition between stellar and DM particles.

Our explanation 
is that 2-body scattering leads to a slow diffusion of stellar particles out of the
dense central regions of galaxies. This is consistent with the simulations of \citet[][$\mu=21.9$]{Revaz2018},
in which quenched dwarf galaxies grow
systematically in size with decreasing $z$, despite their passive evolution.
Indeed, \citet{Revaz2018} hypothesise that this result is due to 2-body scattering. 

We note that hydrodynamical simulations involve gas particles that may also be subject
to mass segregation if their masses differ from those of the DM, but in this case there is a
compounding effect: collisions with DM particles also tend to {\em heat} gas particles
as the kinetic energy associated with velocity perturbations thermalises
\citep[see][for details]{Steinmetz1997}. Disentangling these
effects (mass segregation and gravitational heating) will be challenging, and requires
further study.

Finally, we note that assessing the impact of equipartition on galaxy sizes 
{\em does not} require time-consuming, high-resolution simulations
of large volumes. Since the effect appears limited to haloes/galaxies of relatively low-particle
number it can be gauged by comparing runs in relatively small-volumes that reach the target stellar
mass resolution but vary $\mu$. 2-body scattering may also affect 
velocity dispersion and anisotropy profiles, angular momentum distributions and gas fractions.
These issues will be addressed in a follow-up paper. 


\section*{Acknowledgements}
We thank Adrian Jenkins and Chris Power for useful conversations,
and our referee, Alexander Knebe, for a useful report. ADL is
supported by the Australian Research Council (project Nr. FT160100250).
This work used the DiRAC@Durham facility managed by the Institute for
Computational Cosmology on behalf of the STFC DiRAC HPC Facility
(www.dirac.ac.uk). The equipment was funded by BEIS capital funding
via STFC capital grants ST/K00042X/1, ST/P002293/1, ST/R002371/1 and
ST/S002502/1, Durham University and STFC operations grant
ST/R000832/1. 


\bibliographystyle{mn2e}
\bibliography{paper}

\end{document}